\documentclass[a4paper,11pt]{article}

\usepackage{jcappub}
\usepackage[T1]{fontenc}
\usepackage{multirow}
\usepackage{jabbrv}
\usepackage{bm}
\usepackage{mathrsfs}
\usepackage{amsbsy}
\usepackage{rotating}
\usepackage{enumitem}

\title{Forecast analysis and focal plane optimization for a multi-frequency CMB B-modes polarization experiment: the case of LSPE}

\author[a]{R. Mainini,}
\author[b]{M. Gervasi,}
\author[b]{M. Zannoni,}
\author[c]{L. Lamagna}
\affiliation[a]{Physics Department, University of Milano--Bicocca, \\
Piazza della Scienza 3, I20126, Milano, Italy }
\affiliation[b]{Physics Department, University of Milano--Bicocca and INFN - Milano--Bicocca, \\
Piazza della Scienza 3, I20126, Milano, Italy }
\affiliation[c]{Physics Department, Sapienza University of Rome, \\
Piazzale Aldo Moro 5, I00185, Roma, Italy }

\emailAdd{roberto.mainini@mib.infn.it}

\abstract{
We present an optimization scheme for the focal plane of a multi--frequency 
CMB $B$--modes experiment with a fixed number of detectors 
and apply it to the specific case of LSPE experiment. 
Optimal focal planes are identified on the ground of different figures of merit
defined in terms of the forecasted uncertainty $\sigma_r$ on
the tensor--to-scalar ratio $r$ and the expected map variance from
foreground and instrumental noise residuals.
We then perform a forecast analysis in order to assess the precision
achievable in $B$--modes measurements.

}

\begin{document}
\maketitle
\flushbottom

\section{Introduction}

Detecting $B$--modes of polarization represents a challenge for the next
generation of CMB experiments. The importance of such measurements regards the intimate connection between $B$--modes and the physics of primordial
Universe. As matter of fact $B$--modes were originated from the stochastic background of gravitational
waves expected from a wide class of inflationary models.
The detection of $B$--modes at large angular scales, in addition to being
the definitive confirmation of the inflation theory, offers
the unique opportunity to probe energy scales as high as
$3.3 \times  10^{16} \,r^{1 \over 4}\,$GeV, out of the target of any particle
accelerator (here and in the following $r$ is the tensor--to--scalar 
ratio at the pivot scale $k_*=0.05$ Mpc$^{-1}$
which parametrizes
the amplitude of the primordial gravitational wave signature in the CMB
polarization).

A reliable detection of $B$--modes is however a difficult task.
In addition to being extremely tiny, their signal is buried in the diffuse
galactic foregrounds, namely the polarized synchrotron and thermal dust
emissions, requiring technologically advanced instruments, utmost
sensitivity, unprecedent control of systematic effects, as well as
a wide frequency coverage in order to disentangle or reduce foreground
contaminations.
The strongest constraints to date on the tensor--to--scalar ratio $r$ are
provided by the analisys in \cite{bicep2keck} yielding $r<0.07$ at 95 \%
confidence level.

To further constrain the value of $r$ or, hopefully, achieve the detection
of B-modes, ground based or balloon--borne experiments are being performed or
planned. The LSPE (Large Scale Polarization Explorer) mission,
supported by the Italian Space Agency (ASI), is designed for this endeavor.
It consists of two instruments composed by array of detectors: the
Short Wavelength Instrument for the Polarization Explorer (SWIPE) and
the STRatospheric Italian Polarimeter (STRIP), observing
in different frequency ranges the same sky fraction ($\sim 20-25$\%).
SWIPE is a balloon--borne bolometric instrument which will operate
during the Arctic night (at latitude around 78 degree and for around 15 days)
at three frequency bands centered at 140, 220 and 240 GHz, with an angular resolution of about 1.5 degrees.
STRIP is the ground--based, low frequency module of LSPE. It consists of an 
array of coherent polarimeters which will survey the sky from Teide 
observatory at Tenerife for 1--2 years
in frequency bands centered at 43 and possibly 90 GHz, with an angular resolution of about 0.5 degrees.
Further details on instruments characteristics and informations on LSPE experiment can be found in Table
\ref{noisetab} and \cite{buzzelli,gualtieri,dematteis,bersanelli,aiola,debernardis}.
In its latest proposed configuration, the focal plane of STRIP is
composed by 49 detectors at 43 GHz, plus 4/6 detectors at 90 GHz mainly used as monitors of atmospheric emission,
while SWIPE includes two focal planes each consisting of 3 frequency channels: 140, 220 and 240 GHz with 55, 56
and 52 detectors respectively, for a total of $163 \times 2=326$ detectors.
At this time, due to instrument's geometry and some technological details,
this solution is the favorite one, although there is still room for slight changes.

The aim of this note is the investigation of the optimal focal plane
configurations of the LSPE experiment for primordial $B$--modes measurements.
In the following we therefore consider deviations from the above
configuration (also considering different combination of frequencies)
in order to check whether the performance in constraining $r$
can be improved.
Our criteria to identify the optimal configurations will be the values of
the forecasted uncertainty $\sigma_r$ on the tensor--to--scalar ratio $r$
and/or the expected map variance from foreground and
instrumental noise residuals. We will see that configurations yielding the
lowest variance do not necessarily minimize the uncertainty on $r$.
 A similar but more
sophisticated analysis was performed in \cite{errard} and applied to CMBpol
and COrE satellite experiments while a detailed investigation of the 
capability of LSPE (also in combination with other CMB experiments)
to constrain inflationary parameters will be performed in \cite{maininiLSPE}.

Although specific for LSPE, our results could nevertheless be of interest
for the design of similar experiments. In our analysis we did not consider several sources of systematics not so trivial to be taken into account, such as the non-circularity of the beam pattern or phase-angle uncertainty which can introduce a leakage into detected $B$--modes. Regarding STRIP we did not consider systematic effects coming from the atmospheric fluctuations or long term instabilities, in particular, as suggested in literature \cite{spinelli,battistelli,mainini}, we neglect possible residual polarization of the atmospheric emission.

The plan of the paper is as follows: in Section \ref{FMforecasts} we describe
the Fisher Matrix based method adopted to forecast
the achievable precision on $r$ while in Section \ref{fpopt} our approach
to focal plane optimization and our findings are presented.
Results are first given by considering only the high frequency instruments
SWIPE and $B$--modes measurements. We then consider the combination SWIPE+STRIP
and include the $E$--modes in the analysis. Effects of detector noise and
frequency channel correlations on results are also discussed.
Finally, Section \ref{conclusions} is devoted to conclusions.
Appendices collect some material needed to implement our analysis:
in Appendix \ref{foregroundremoval} we describe the adopted method for
estimating galactic  foreground (and instrumental noise) residuals
to be included in the Fisher Matrix forecasting analysis.
Foregrounds are modeled according to recent findings of Planck collaboration
and detailed in Appendix \ref{foremodels} while characteristics of LSPE
instruments are summarized  in Appendix \ref{noise}. The polarization angular
power spectra covariance matrix entering the Fisher Matrix is given in
Appendix \ref{covariance}.

\vfill
\eject

\section{Fisher Matrix forecasts}
\label{FMforecasts}
In order to estimate the precision achievable on $r$ we make use of the
Fisher Matrix (FM) approach.
Let $C_l^X$ and ${\rm R}_l^X$ ($X=E,B$) denote
the angular power
spectra of CMB polarization and residual non--cosmic signal (foregrounds and
instrumental noise) respectively.
We then define the binned power spectra in multipole bands $b$
of width $\Delta l$:
\begin{equation}
{\rm D}_b^X= \frac{1}{\Delta l}\sum_{l\in b} {\rm D}_l^X
\label{binnedspectra}
\end{equation}
where
\begin{equation}
{\rm D}_l^X=\frac{l(l+1)}{2\pi}\left( C_l^X
+ {\rm R}_l^X \right)
\label{spectra}
\end{equation}
The Fisher Matrix then reads:
\begin{eqnarray}
\nonumber
F_{ij} &=& \sum _X\sum_b F_{ij,b}^X + \frac{\delta_{ij}}{\sigma_P^2(p_i)} \\
&=& \sum _X\sum_b
\frac{\partial {\rm D}_b^{X}}{\partial p_i}
\left[{\bf D}_b^{-1}\right]_{XX}
\frac{\partial {\rm D}_b^X}{\partial p_j}
+ \frac{\delta_{ij}}{\sigma_P^2(p_i)}
\label{FM}
\end{eqnarray}
Here $\sigma_P(p_i)$ denotes a possible Gaussian prior on the
model parameter $p_i$ and the derivatives are evaluated
at the fiducial model values.
The covariance matrix  of the binned power spectra, ${\bf D}_b$,
is given in Appendix \ref{covariance}.
The inverse of FM then gives the covariance matrix $C_{ij}$ of the model
parameters, the diagonal
elements of which represent the lowest variance $\sigma^2(p_i)$ one can
achieve on the parameter $p_i$.

The limited sky fraction observed by LSPE ($f_{sky}
\simeq 0.2$) places a limit on the minimum resolution  $\Delta_l$
of the angular power spectra under which different multipoles become
correlated. This is approximately given by
$\Delta_l=\pi/\Theta$ ($\Theta^2$ being
the survey area), thus, in our analysis we consider multipole bins
of width $\Delta_l=3$.
Since LSPE will measure polarization, the only
relevant foregrounds included in the analysis are polarized dust
and syncrothron emissions which we model according to the recent findings
of the Planck collaboration (see Appendix \ref{foremodels}).

In order to perform our FM analysis we need to estimate the residuals ${\rm R}_l^X$,
arising after some foreground (and instrumental noise) cleaning procedure.
To this aim, here, we adopt the method proposed by \cite{tegmark} and detailed
in Appendix \ref{foregroundremoval}.

As a base cosmological model we assume a flat $\Lambda$CDM cosmology with
tensor perturbations described by the following
seven parameters:
$$
\{ \Omega_b h^2, \Omega_c h^2, \theta, \tau, n_s, \ln(10^{10} A_s), r \}
$$
where $\Omega_b$ are $\Omega_c$ are the baryon
and cold dark matter density parameters respectively, $h$ is the
dimensionless Hubble parameter, $\theta$ is the angle subtended by the
sound horizon at recombination, $\tau$ is the optical depth to reionization,
$n_s$ is the scalar spectral index, $A_s$ is the amplitute of scalar fluctuations
at a pivot scale $k^*=0.05$ Mpc$^{-1}$ and $r$ is the tensor to
scalar ratio.
Fiducial values of the six standard cosmological parameters
$\{ \Omega_b h^2, \Omega_c h^2, \theta, \tau, n_s, \ln(10^{10} A_s) \}$
are  set to the best fit values estimated by Planck Collaboration
2015 \cite{planckXIII2015} (Planck $TT,EE,TE+{\rm low} P$ data) while we assume
a fiducial $r=0.05$ all through the paper with the exception of section
\ref{Emodes}
where the range $0.01 \leq r \leq 0.1$ is considered.

Since LSPE measurements will be limited to the first $\sim 140$ multipoles
of polarization spectra, strong constraints are expected only on $r$
and $\tau$. Therefore, when applying the FM approach all the cosmological
parameters will be kept fixed to
their fiducial values except for:

i) $r$ if only $B$--modes are included in the analysis,

ii) $r$, $\tau$ and $\ln(10^{10} A_s)$ (the latter being strongly correlated with
$\tau$) if both $E$-- and $B$--modes are included.

Moreover, amplitudes ${\cal A}_{D(S)}^X$ and spectral indices $\beta_{D(S)}$
of dust ($D$) and syncrothron ($S$) spectra
will be included in the analysis as nuisance parameters and marginalized over.
Foreground models and adopted priors are detailed in Appendix \ref{foremodels}
while Appendix \ref{noise} describes the characterisics of LSPE instruments.

\begin{figure}[]
\begin{center}
\includegraphics[angle=-90, width=1.0\textwidth]{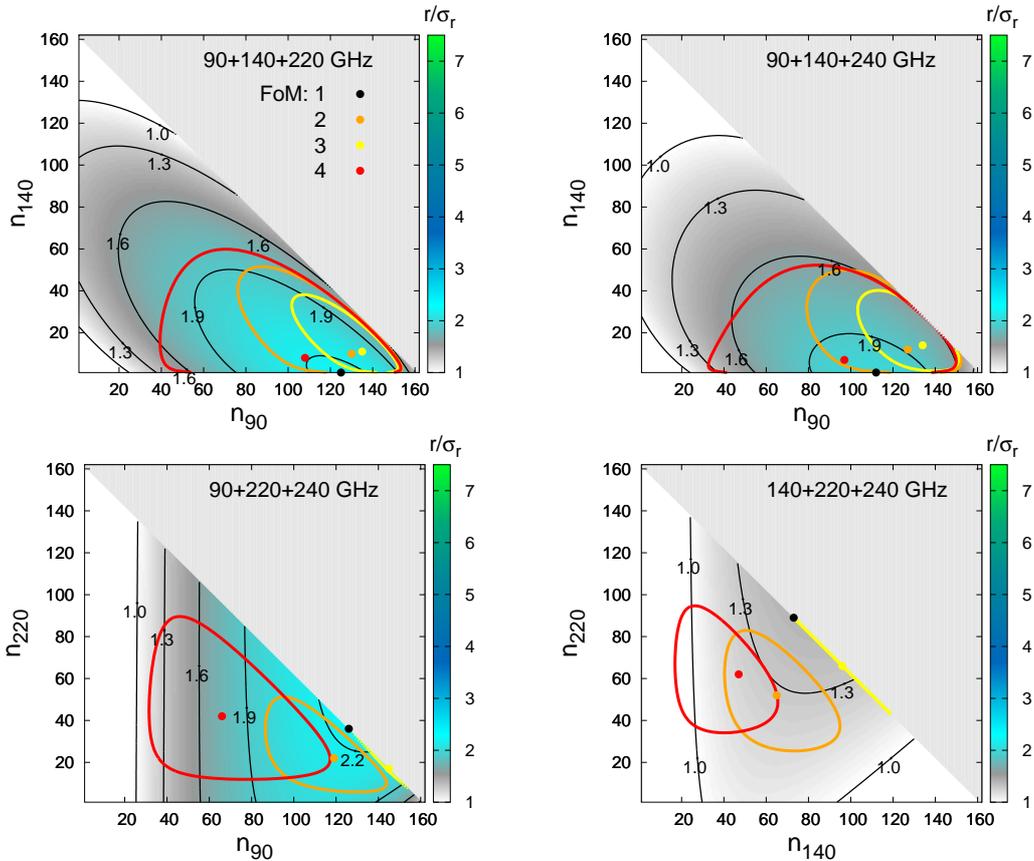}
\caption{Contour levels for $r/\sigma_r \geq 1$ in steps of 0.3 for different
combinations of SWIPE frequency channels.
Black, orange, yellow and red dots indicate configurations satisfying
FoM 1, 2, 3 and 4 respectively (see text). Orange, red and
yellow curves surround the configurations for which the variances,
$\sigma^2_{\rm tot}$, $\sigma^2_{\rm for}$ and $ \sigma^2_{\rm det}$, deviate
by no more than 5$\%$ from their minimum values.}
\label{confronto}
\end{center}
\end{figure}

\begin{table}
\begin{center}
\renewcommand\arraystretch{1.2}
\begin{tabular}{|ccccc|ccc|}
\hline
\multicolumn{8}{|c|}{90+140+220 GHz} \\
\hline
  FoM    & $r/\sigma_r$ & $\sigma_{\rm tot}^2/\mu {\rm K}^2 $ & $\sigma_{\rm det}^2 /\mu {\rm K}^2$ & $\sigma_{\rm for}^2 /\mu {\rm K}^2$ & $n_{90}$  & $n_{140}$ & $n_{220}$ \\
\hline
  1  &  {\bf 2.2400} &   0.0814 &   0.0234 &   0.0580 &     125 &   1 & 37  \\
  2  &   2.1325 &   {\bf 0.0783} &   0.0213 &   0.0570 &     130 &  10 & 23  \\
  3  &   2.0569 &   0.0784 &   {\bf 0.0211} &   0.0573 &     135 &  11 & 17  \\
  4  &   2.1937 &   0.0802 &   0.0235 &   {\bf 0.0566} &     108 &   8 & 47  \\
\hline
\multicolumn{8}{c}{} \\
\hline
\multicolumn{8}{|c|}{90+140+240 GHz} \\
\hline
FoM      & $r/\sigma_r$ & $\sigma_{\rm tot}^2 /\mu {\rm K}^2$ & $\sigma_{\rm det}^2 /\mu {\rm K}^2$ & $\sigma_{\rm for}^2 /\mu {\rm K}^2$ & $n_{90}$  & $n_{140}$ & $n_{240}$ \\
\hline
  1  & {\bf 2.0322}  &   0.0844  &   0.0263   &  0.0581    &   112  &   1  & 50   \\
  2  & 1.8790  &   {\bf 0.0797}  &   0.0227   &  0.0570    &   127  &  12  & 24   \\
  3  & 1.7898   &  0.0800  &   {\bf 0.0225}    & 0.0575    &   134  &  14  & 15   \\
  4  & 1.9778   &  0.0828  &   0.0265    & {\bf 0.0564}    &   97   &   7  & 59  \\
\hline
\multicolumn{8}{c}{} \\
\hline
\multicolumn{8}{|c|}{90+220+240 GHz} \\
\hline
FoM      & $r/\sigma_r$ & $\sigma_{\rm tot}^2 /\mu {\rm K}^2$ & $\sigma_{\rm det}^2 /\mu {\rm K}^2$ & $\sigma_{\rm for}^2 /\mu {\rm K}^2$ & $n_{90}$  & $n_{220}$ & $n_{240}$ \\
\hline
  1 & {\bf   2.2368}   &   0.0888   &   0.0257  &    0.0632   &      126  &   36  &  1   \\
  2  & 2.1636   &  {\bf   0.0763}   &   0.0276   &   0.0487   &      119  &   22  &    22    \\
  3  & 2.0991   &   0.0885   &  {\bf   0.0244}   &   0.0641    &     145  &   17  &  1   \\
  4  & 1.7552    &  0.0873   &   0.0411   &   {\bf  0.0462}    &     66   &  42   &  55    \\
\hline
\multicolumn{8}{c}{} \\
\hline
\multicolumn{8}{|c|}{140+220+240 GHz} \\
\hline
FoM      & $r/\sigma_r$ & $\sigma_{\rm tot}^2 /\mu {\rm K}^2$ & $\sigma_{\rm det}^2 /\mu {\rm K}^2$ & $\sigma_{\rm for}^2 /\mu {\rm K}^2$ & $n_{140}$  & $n_{220}$ & $n_{240}$ \\
\hline
  1 &  {\bf  1.3980}  &    0.1662 &     0.0400 &     0.1262  &      73 &    89 &    1 \\
  2 &  1.2845  &    {\bf  0.1159} &     0.0509 &     0.0650  &      65 &    52 &    46 \\
  3 &  1.3371  &    0.1657 &     {\bf  0.0384} &     0.1273  &      96 &    66 &    1 \\
  4 &  1.1202  &    0.1263 &     0.0644 &     {\bf  0.0619}  &      34 &    63 &    66  \\
\hline
\end{tabular}
\caption{Values of $r/\sigma_r$,  variances and the number of detectors
corresponding to the
configurations satisfying FoM 1, 2, 3 and 4 for different combinations
of SWIPE frequency channels.}
\label{confrontoval}
\end{center}
\end{table}

It is worth noticing that the simplified procedure here adopted does not
address a number of issues. For partial sky experiments the number of modes
of multipole $l$ available for the analysis decreases with the sky coverage.
As usual, to capture this effect we simply assume that
the number of independent modes of a given multipole is approximately
reduced by a factor of $f_{sky}$, i.e. $(2l+1) \rightarrow (2l+1)f_{sky}$.
This is taken into account in the definition of ${\bf D}_b$.
However, in case of polarization, partial sky coverage causes a leakage
of $E$--modes into $B$--modes (neglected in our analysis) which can alter
the scaling with $f_{sky}$ \cite{amarie} other than add complications to the
detection of primordial gravitational waves. Another issue concerns the
$B$--modes signal induced by weak lensing  which is a limiting factor
in measuring $B$--modes of primordial origin.
We assume no removal of lensing contamination
and treat it as a Gaussian noise for simplicity. However,
lensing is not important at low multipoles if the tensor to scalar ratio
is large or when the residual noise is larger than the lensing signal.
For simplicity we also assume foregrounds to have a Gaussian probability
distribution although, generally, they do not. We also neglect correlations
among galactic dust and syncrothron recently measured by \cite{planckXXII,choi}.

\section{Focal plane optimization}
\label{fpopt}
\subsection{High frequencies}
\label{high}
In this Section we only consider the high frequency instrument (SWIPE)
and $B$--modes measurements so that the set of parameters to vary in
FM analysis is ${\bf p}=\{ r,{\cal A}_S^B,{\cal A}_D^B,\beta_S,\beta_D \}$.

We assume each of the two (identical) focal planes to consist of 3
frequency channels
($\nu_i$, $i=1,2,3$) each with $n_{\nu_i}$ detectors for a total
of $n_{tot}=163$ detectors. Further, all the detectors are supposed to have
the same size so that different combinations occupy the same area.

Next we consider all the configurations
obtained by varing the $n_{\nu_i}$'s such that $n_{\nu_1}+n_{\nu_2}+n_{\nu_3}=163$
and apply the cleaning procedure of Appendix \ref{foregroundremoval} and the
FM formalism to each of them. We then define the following
Figures of Merit (FoM):
\begin{description}[noitemsep]
\item [FoM 1] maximum signal to noise ratio $r/\sigma_r$
\item [FoM 2] minimum variance, $\sigma^2_{\rm tot}$, from residual
foregrounds and detector noise
\item [FoM 3] minimum variance, $\sigma^2_{\rm det}$, from residual
detector noise
\item [FoM 4] minimum variance, $\sigma^2_{\rm for}$, from residual
foregrounds
\end{description}
where:
$$
\sigma^2_{\rm tot}=\sum_b {\rm R}_b^B \sum_{l \in b} \frac{2l+1}{4\pi}W_l
$$
(${\rm R}_b^B$ is the binned angular power spectrum of residual foregrounds
and detector noise, $W_l$ is the beam window function) and similar expressions
hold for $\sigma^2_{\rm for}$ and $\sigma^2_{\rm det}$.
Finally, we select and compare the configurations that satisfy the above FoM's.

\medskip
We start by considering all the combinations of 3 frequencies
chosen among 90, 140, 220 and 240 GHz.
Results are displayed in Figure \ref{confronto} which shows:
i) contour levels for $r/\sigma_r \geq 1$ in steps of 0.3,
ii) configurations satisfying FoM 1, 2, 3 and 4 (black, orange, yellow and
red dots repectively; orange, red and
yellow curves surround the configurations for which the variances,
$\sigma^2_{\rm tot}$, $\sigma^2_{\rm for}$ and $ \sigma^2_{\rm det}$, deviate
by no more than 5$\%$ from their minimum values).
Table \ref{confrontoval} summarizes the values of $r/\sigma_r$,
the variances and the number of detectors corresponding to the
configurations in ii).

\begin{figure}[]
\begin{center}
\includegraphics[angle=-90, width=1\textwidth]{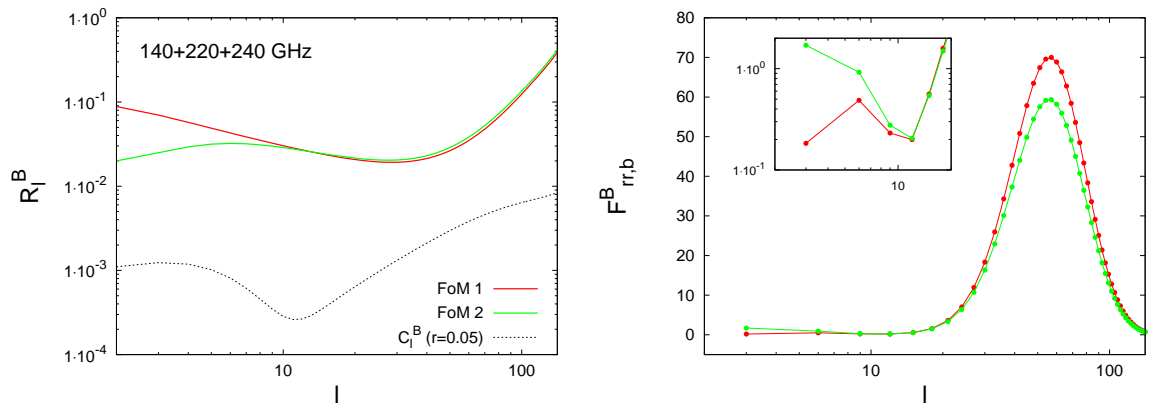}
\caption{{\it Left panel}: angular spectra of residual non--cosmic
signal in the cases FoM 1 and FoM 2 for the combination SWIPE
$140+220+240$ GHz.
{\it Right panel}: contributions to Fisher Matrix in the two cases.}
\label{fish}
\end{center}
\end{figure}

We first note that the combination $140+220+240$ GHz provides the worst results.
This is due to the absence of the frequency channel at 90 GHz for which foregrounds
are minimal and the instrumental noise is low (see Table \ref{noisetab}).
Indeed, when present,
the 90 GHz channel is assigned with the majority of detectors.
We also observe that requiring the variance to be minimal does not guarantee the best
precision on $r$, although in some cases differences are only marginal.
This can be understood by inspection of Figure \ref{fish} where, for illustrative
purpose, we consider the cases FoM 1 and FoM 2 for the
combination $140+220+240$ GHz.
The left panel compares the spectra of residual non--cosmic signal,
${\rm R}_l^B$, in the two cases. FoM 2 allows for a better foreground cleaning
at low multipoles ($l \lesssim 10$) wich results in a lower variance $\sigma^2_{\rm tot}$.
Nevertheless, the better precision $\sigma_r$
($ \simeq 1/\sqrt {F_{rr}}$ for negligible correlations among $r$ and  foreground
parameters) achievable in the case FoM 1, is to
be ascribed to the (sligth) lower ${\rm R}_l^B$ at $l \gtrsim 20$ which makes
the inverse covariance matrix ${\bf D}^{-1}_b$ to increase. This can be inferred from
the right panel of Figure \ref{fish} showing that  most of the contributions
$F_{rr,b}^B \propto {\bf D}^{-1}_b $, to the Fisher Matrix come from multipoles in
the range $40 \lesssim l \lesssim 80$.

In the following we consider only the combination of frequencies having the best (worst)
performance, namely $90+140+220$ GHz ($140+220+240$ GHz). Although the 90+220+240 GHz
combination performs similarly to $90+140+220$ GHz, we prefer the latter since
configurations satisfying our FoM's are quite close each other, all laying
in the region with the higher $r/\sigma_r$ (see Fig. \ref{confronto}).

\subsubsection{Reducing the detector noise}
\label{noisered}
\begin{figure}[]
\begin{center}
\includegraphics[angle=-90, width=1.\textwidth]{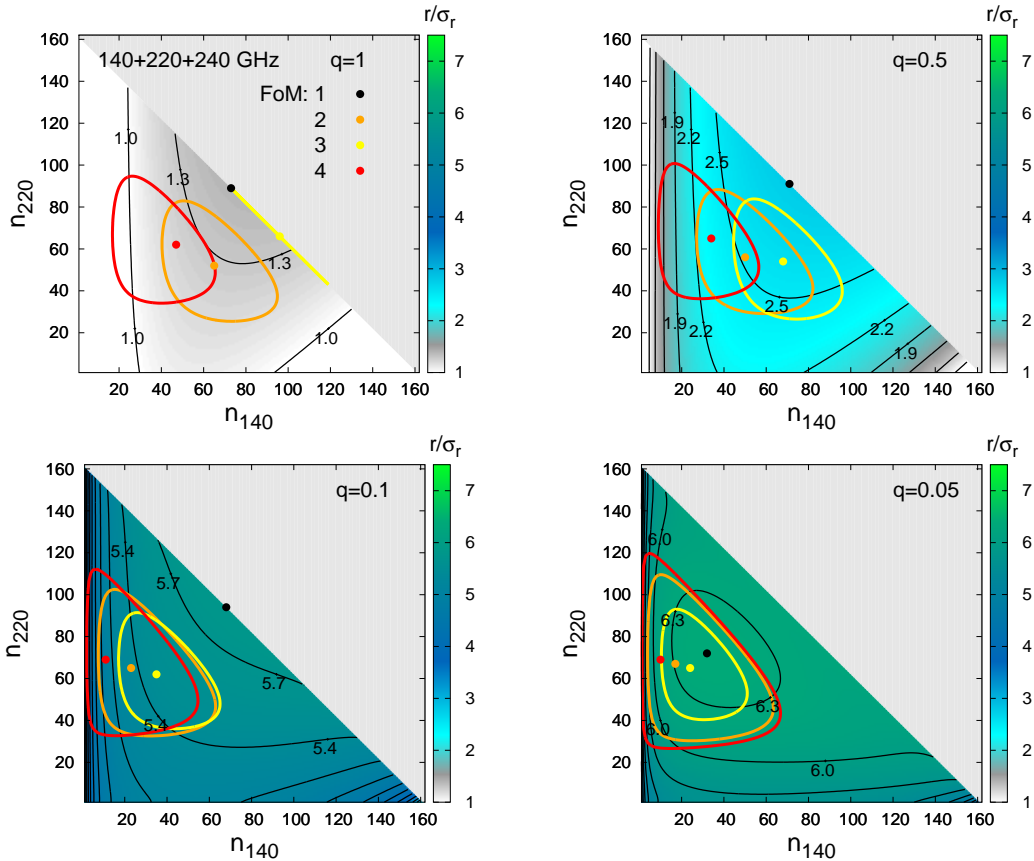}
\caption{Changes in the features of focal plane configurations
for SWIPE $140+220+240$ GHz
as the detector noise
is reduced by a factor $q=0.5, 0.1, 0.05$. Lines and colors are like in Figure
\ref{confronto}.}
\label{c140}
\end{center}
\end{figure}
\begin{figure}[]
\begin{center}
\includegraphics[angle=-90, width=1.\textwidth]{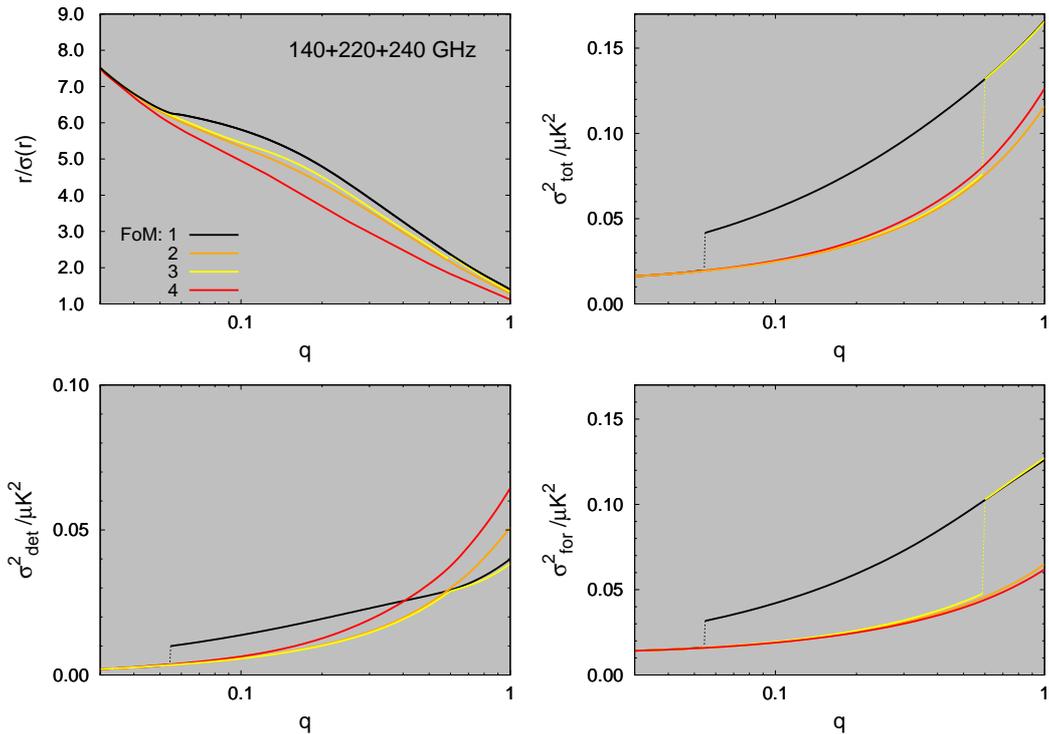}
\caption{$r/\sigma_r$ and $\sigma^2_x$ ($x={\rm tot,det,for}$)
as a function of $q$ for FoM 1--4 and SWIPE $140+220+240$ GHz.}
\label{u140}
\end{center}
\end{figure}

\begin{figure}[]
\begin{center}
\includegraphics[angle=-90, width=1.\textwidth]{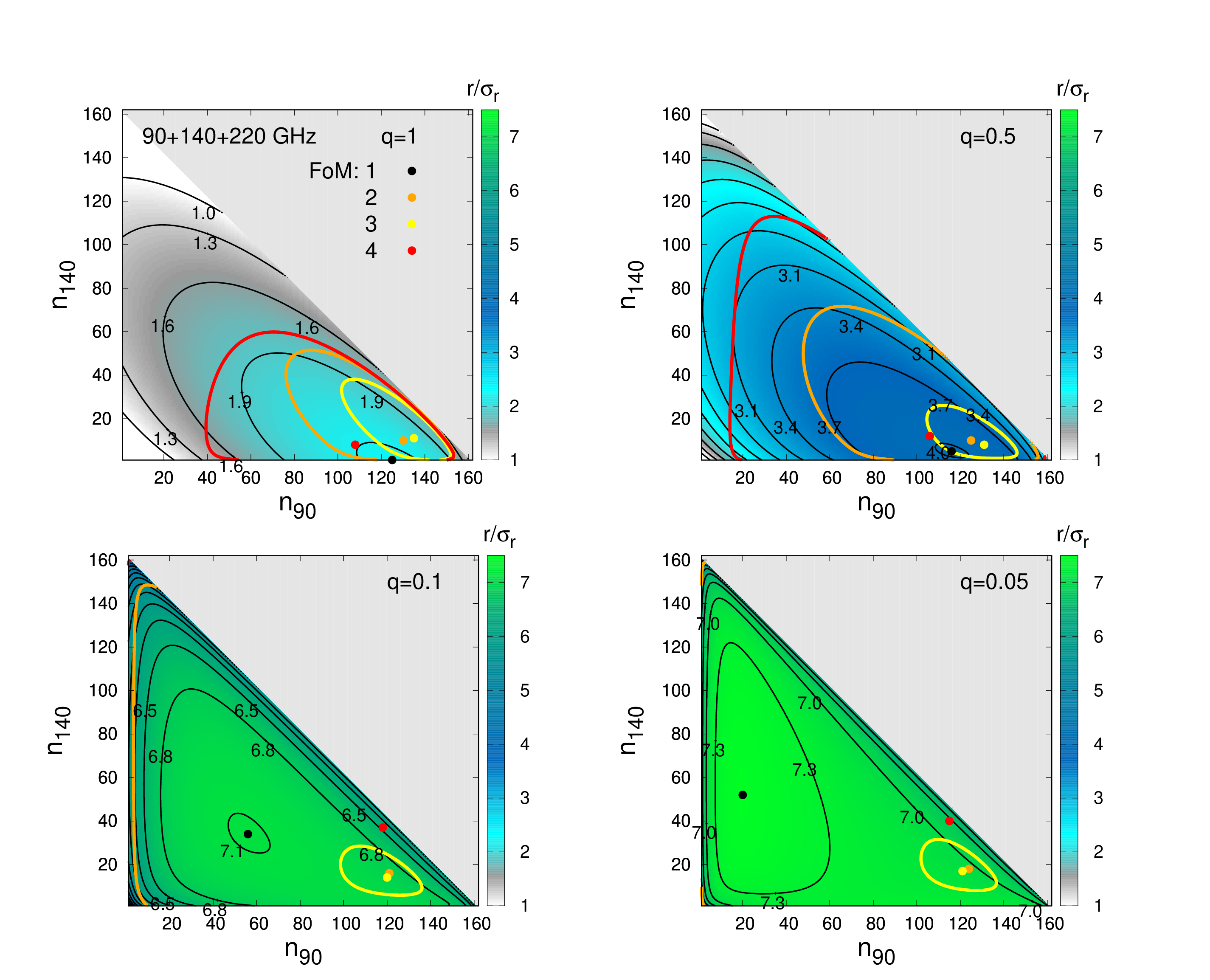}
\caption{As Figure \ref{c140} but for SWIPE $90+140+220$ GHz.}
\label{c90}
\end{center}
\end{figure}
\begin{figure}[]
\begin{center}
\includegraphics[angle=-90, width=1.\textwidth]{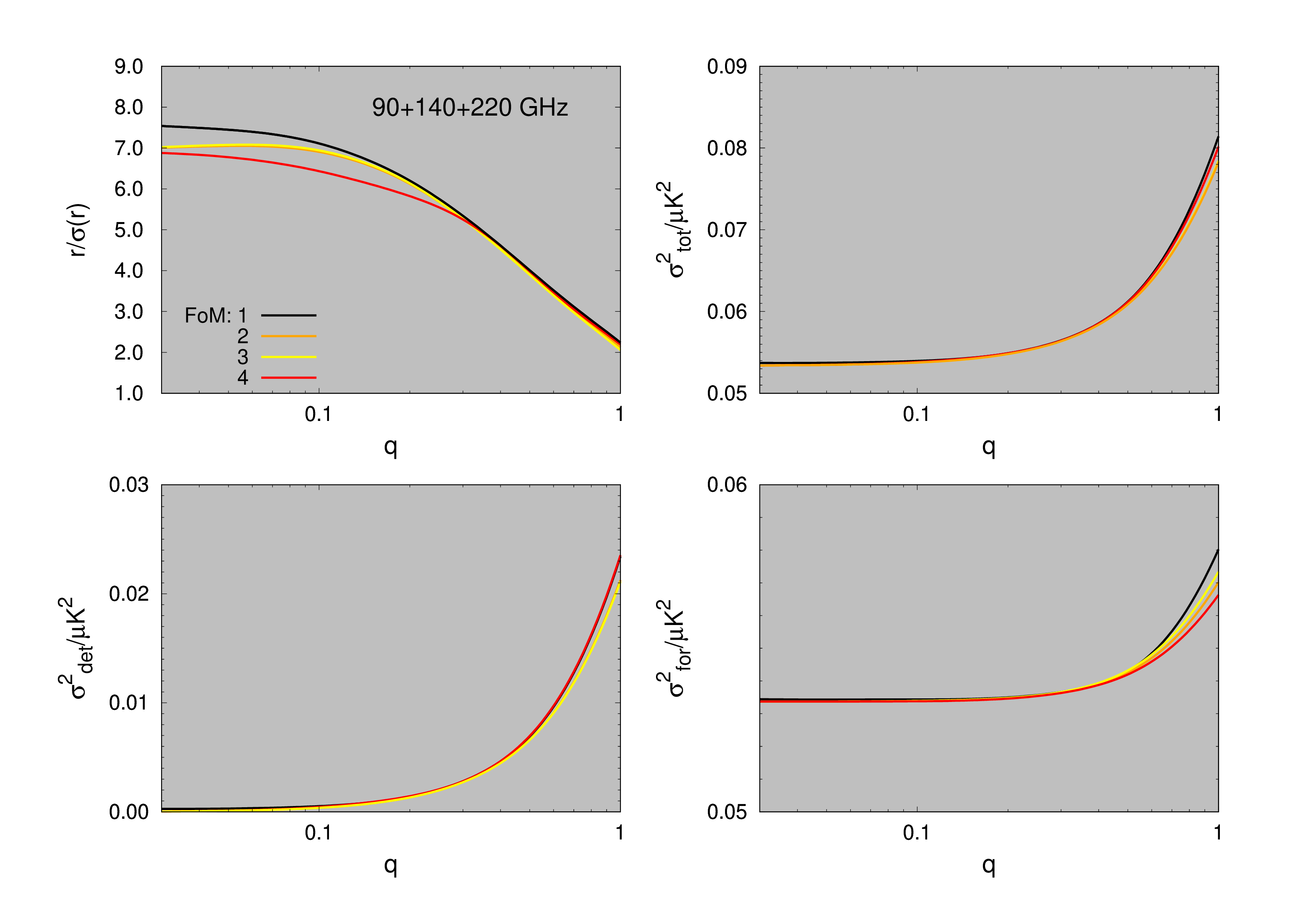}
\caption{As Figure \ref{u140} but for SWIPE $90+140+220$ GHz.}
\label{u90}
\end{center}
\end{figure}

\begin{figure}[]
\begin{center}
\includegraphics[angle=-90, width=1.\textwidth]{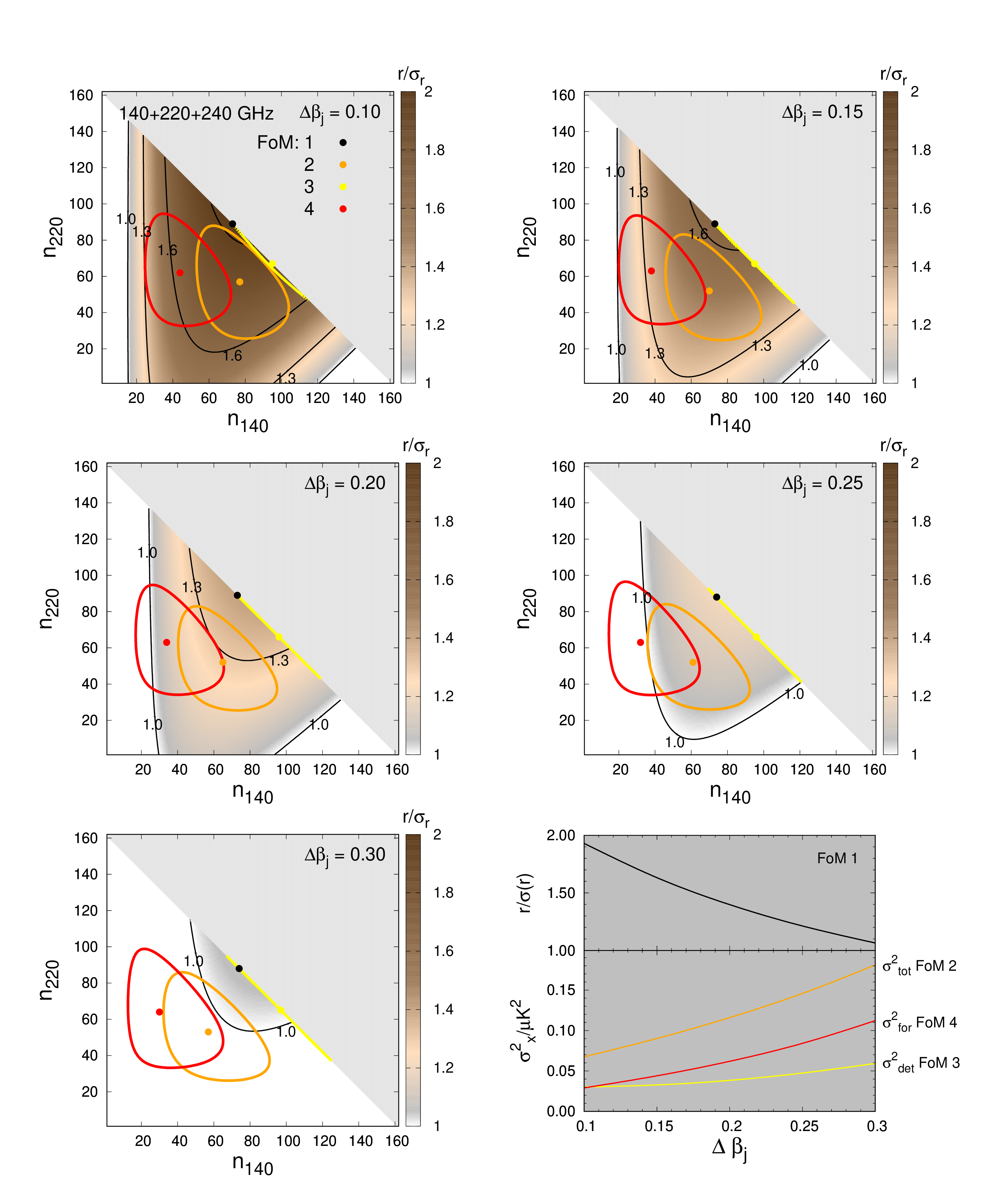}
\vskip 0.8truecm
\caption{The effect of frequency channel correlations on focal
plane configurations for SWIPE $140+220+240$ GHz.
Apart the different contour scale for $r/\sigma_r$,
lines and colors are as in previous Figures. The right bottom panel
shows $r/\sigma_r$, $\sigma^2_{\rm tot}$, $\sigma^2_{\rm det}$ and
$\sigma^2_{\rm for}$
for FoM 1, 2, 3 and 4 respectively as a function of the
{\it frequency coherence}
$\xi_j \approx 1/\sqrt2 \Delta\beta_j $ ($j=S,D$).}
\label{dalfa}
\end{center}
\end{figure}

In this section we inspect the effect of reducing the instrumental
noise. To this aim the sensitivities $\sigma_{pix}$ of each channel
(see Table \ref{noisetab})
are simultaneously reduced by a common factor $q$.
Let us first consider the combination $140+220+240$ GHz. Figure \ref{c140}
illustrates how the configuration's features change when the detector noise
is reduced by $q=0.5,0.1,0.05$ while Figure \ref{u140} shows
the values of $r/\sigma_r$ and $\sigma^2_x$ ($x={\rm tot,det,for}$) obtained
by the configurations satisfying our FoM's as a function of $q$ in the range
$q=0.01-1$.

Focusing first on FoM 1 and Figure \ref{u140}, we can observe
a discontinuity in the behavior of $\sigma^2_x$ (black curves) to occur around
$q\simeq 0.055$ which corresponds to a transition of the configuration FoM 1
from $\{n_{140},n_{220},n_{220}\} \simeq \{67,95,1\}$ to $\{37,73,53\}$ (see Figure
\ref{c140}, black dots).
To understand this behavior, let us remember that most of the contributions to
$r/\sigma_r$ arise from multipoles in the range $l=40-80$ (see section
\ref{high} and Figure \ref{fish}) where the detector noise starts to
increase above the foreground level.
Contributions to $r/\sigma_r$ are then suppressed by growing noise
at larger $l$.
Therefore, in order to maximize $r/\sigma_r$, the optimization procedure
tends to limit the instrumental noise effect omitting noisier channels, i.e.
the 240 GHz channel. On the other hand, as the noise is made smaller,
foregrounds become more and more relevant in the full range of multipoles
considered so that they should be efficiently removed. To this aim,
the 240 GHz channel is also required. Referring to Figure \ref{c140},
once the noise level decreases sufficiently,
a new local maximum for $r/\sigma_r$ forms around $\{37,73,53\}$ becoming
absolute for $q < 0.055$.
Something similar happens in the case of FoM 3. Around $q \simeq 0.6$ a
transition occur from  $\{88,74,1\}$ to $\{72,53,38\}$ causing discontinuities
in  $r/\sigma_r$, $\sigma^2_{\rm tot}$ and $\sigma^2_{\rm for}$ (yellow line in
Figure \ref{u140}).
Figures \ref{c90} and \ref{u90} are similar to Figures \ref{c140} and \ref{u140}
but for the combination $90+140+220$ GHz.

It is however worth mentioning that in order to improve the detector's
sensitivity,
$\sigma_{pix}$, a longer observation time, $t_{obs}$, is required.
According to $\sigma_{pix} \propto 1/\sqrt{t_{obs}}$, already in the case $q=0.5$
we should increase $t_{obs}$ by a factor of 4 which
could be somewhat problematic given the technical constraints of the SWIPE
payload. On the other hand, the noise reduction is also
limited by the level of systematics.

\begin{figure}[]
\begin{center}
\includegraphics[angle=-90, width=0.5\textwidth]{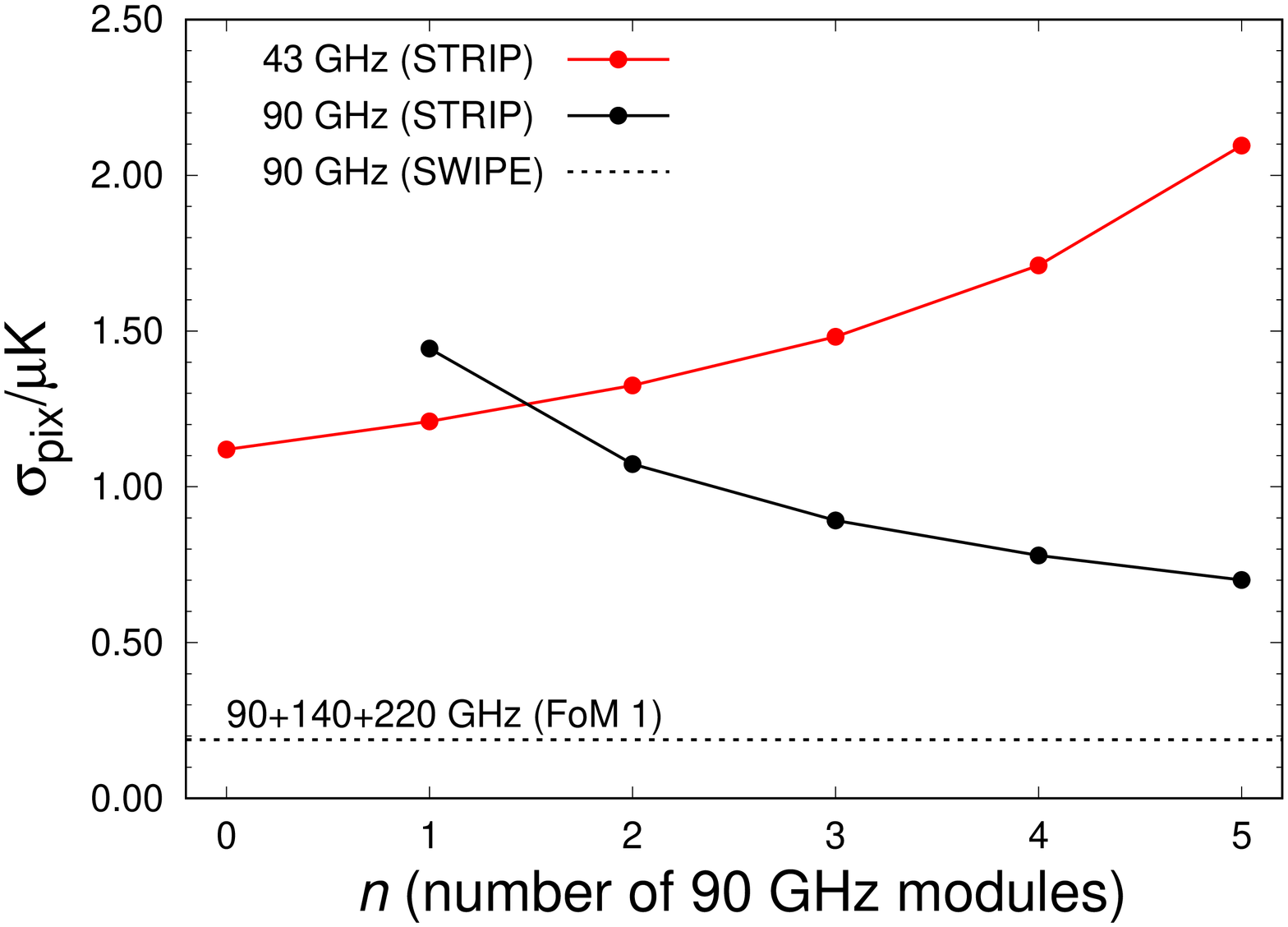}\hfil
\includegraphics[angle=-90, width=0.5\textwidth]{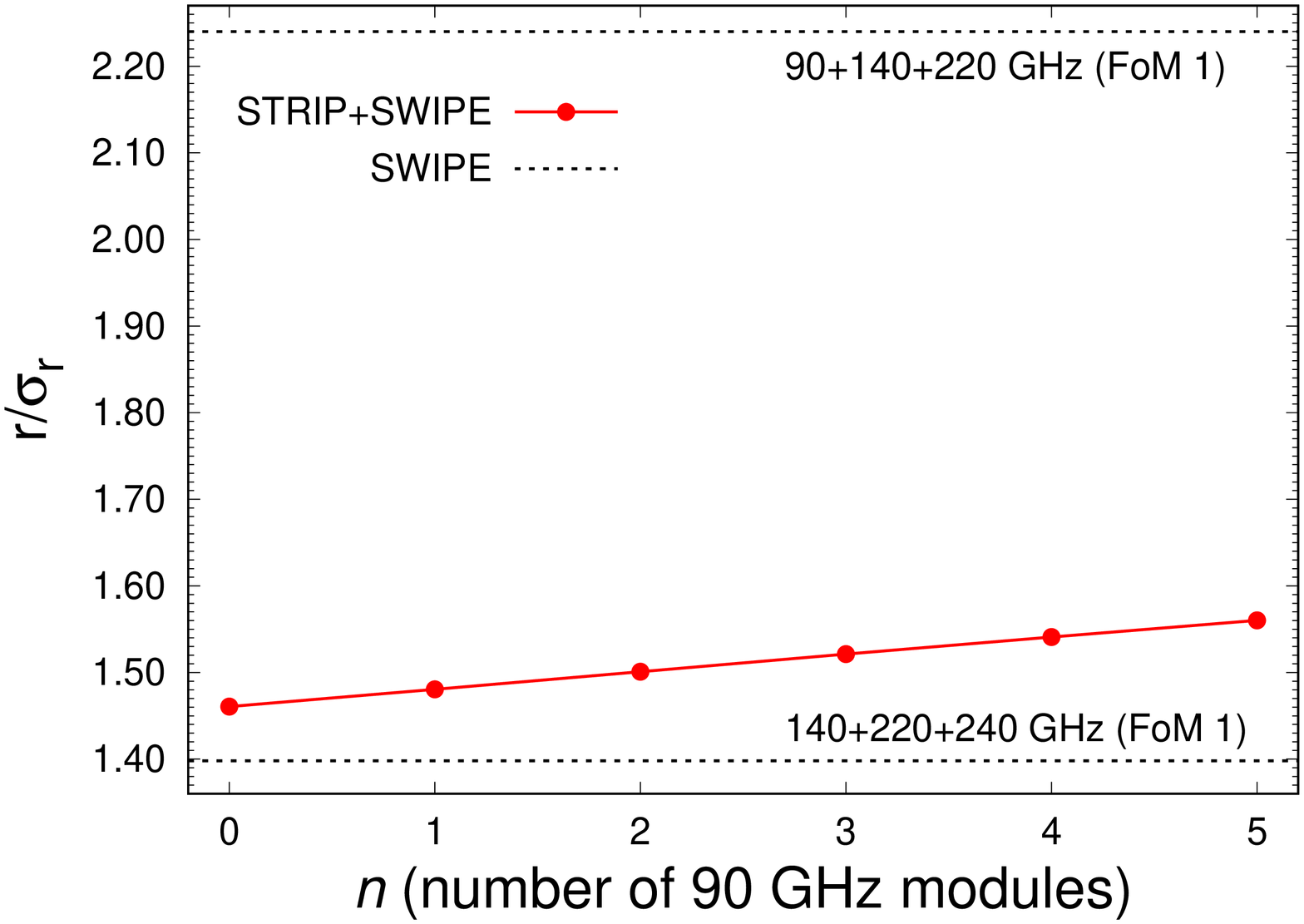}
\caption{{\it Left panel}: sensitivity of STRIP 43 GHz and 90 GHz
channels as a function of $n$, the number of 90 GHz modules in the focal plane
(see text). {\it Right panel}: $r/\sigma_r$ as a function of $n$ for
the STRIP(43+90 GHZ)+SWIPE(140+220+240 GHz) combination.}
\label{low+high}
\end{center}
\end{figure}

\begin{figure}[]
\begin{center}
\includegraphics[angle=-90, width=1.\textwidth]{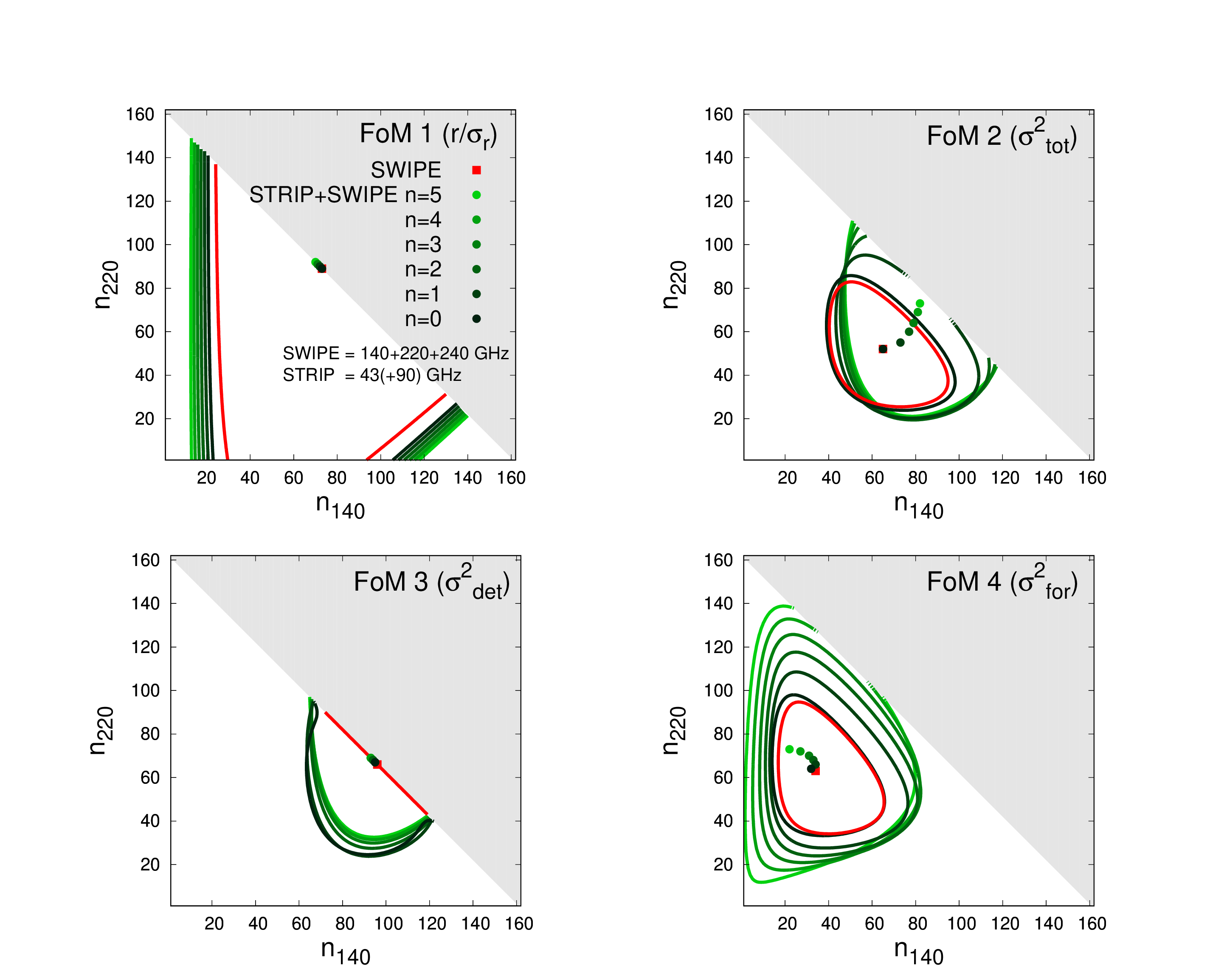}
\caption{STRIP+SWIPE configurations satisfying
FoM 1--4 for $n=0,1,2,3,4,5$ (green dots).
Contours enclose configurations for which  $r/\sigma_r > 1$
(top left panel) or $\sigma^2_{\rm tot}$, $\sigma^2_{\rm for}$ and
$\sigma^2_{\rm det}$ deviate by no more than 5$\%$ from their minimum values
(top right, bottom right and bottom left panel
respectively).
Results expected from SWIPE only (red) are also shown for comparison.}
\label{mod140}
\end{center}
\end{figure}

\begin{figure}[]
\begin{center}
\includegraphics[angle=-90, width=1.\textwidth]{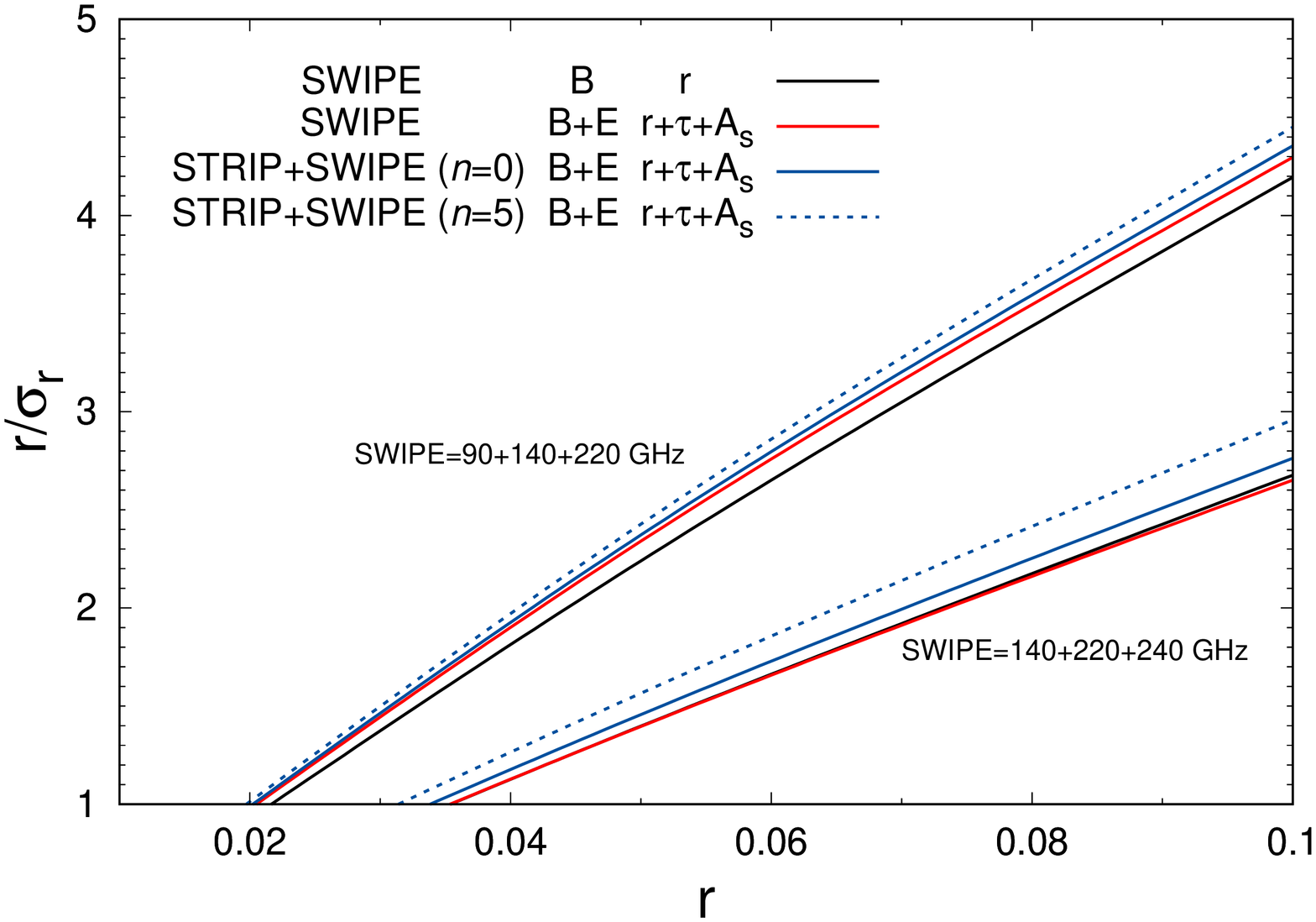}
\caption{$r/\sigma_r$ as a function of $r$ for SWIPE ($B$--modes only),
SWIPE ($B+E$--modes) and SWIPE+STRIP ($B+E$--modes).
Results are given for FoM 1 and
different frequency channel combinations.}
\label{sn}
\end{center}
\end{figure}

\subsubsection{Frequency channel correlations}
As detailed in Appendix \ref{foremodels} and according to
\cite{tegmark98,tegmark} correlations among different
frequency channels are taken into account
by specifying the {\it frequency coherence}
$\xi_j \approx 1/\sqrt2 \Delta\beta_j $ ($j=S,D$) which quantifies
spectral variations
of foregrounds across the sky, $\Delta\beta_j$ being the dispersion of the
foreground spectral index $\beta_j$ through the sky. Two extreme cases
corresponding to no correlation and perfect correlation then arise
in the limits $\xi \rightarrow 0$ and $\xi \rightarrow \infty$ respectively.

In this section we consider deviations from our fiducial value
$\Delta\beta_j = 0.2$ which is consistent with findings by
\cite{planckXXII,fuskeland,kra}.
Results are summarized in Figure \ref{dalfa} for the combination
$140+220+240$ GHz (different combinations giving similar qualitative results).
We note that configurations satisfying our FoM's are not substantially
affected by the value of $\Delta\beta_j$ (or $\xi_j$). Nevertheless, as expected,
$r/\sigma_r$ ($\sigma^2_x$) decreases (increases) as channels become
less correlated (see right bottom panel of Figure \ref{dalfa}).

\subsection{High+Low frequencies}
Let us now consider the high frequency combination $140+220+240$ GHz and
extend our analysis including the low frequency
ground--based instrument (STRIP), observing at 43 and (possibly) at 90 GHz.

In its basic configuration
the STRIP focal plane is supposed to consist of an array of 49 polarimeters at
43 GHz accomodated in 7 modules each including 7 detectors plus
a smaller array of 6 polarimeters at 90 GHz mainly used as monitors of 
atmospheric emission.

Motivated by results of section \ref{high} according to which
the presence of a 90 GHz channel seems to substantially improve the signal 
to noise ratio $r/\sigma_r$,
we also consider the possibility to replace $n$ modules at 43 GHz
with an equal number of 90 GHz modules each accomodating 30 detectors
(here we are supposing that the detector's dimensions
simply scale as the wavelenght so that 30 detectors at 90 GHz with resolution
$\theta_{FWHM}=15^\prime$  or
7 detectors at 43 GHz with resolution $\theta_{FWHM}=30^\prime$ occupy 
the same area).
A generic focal plane, therefore, will consist of $7-n$ modules at 43 GHz
plus $n$ modules at 90 GHz for a total of $(7-n) \times 7$ and $n \times 30$
detectors respectively (plus the 6 detectors at 90 GHz  with resolution 
$\theta_{FWHM}=30^\prime$ devoted to atmospheric emission).

For $n$ fixed, we then proceed as in section \ref{high}  to find the
configurations satisfying our FoM's.
The left panel of Figure \ref{low+high} shows the sensitivities $\sigma_{pix}$
(rescaled to $1^\circ$ pixel)
of 43 and 90 GHz low frequency channels as a function of $n$, the number
of 90 GHz modules in the focal plane.
For comparison the sensitivity of the 90 GHz channel of SWIPE
combination $90+140+220$ GHz (FoM 1, i.e. 125 detectors $\times$ 3 modes) 
is also shown.
As clearly visible, this latter value is always far below the sensitivity of
the STRIP 90 GHz channel meaning that, although the addition of
low frequencies somewhat improve $r/\sigma_r$, this is substantially lower
than the value expected from SWIPE $90+140+220$ GHz (FoM 1)
(see right panel of Figure \ref{low+high}).
Figure \ref{mod140} then displays configurations which satisfy
FoM 1--4 for $n=0,1,2,3,4,5$ (green dots). In the same Figure,
contours enclose configurations for which  $r/\sigma_r > 1$
(top left panel) or $\sigma^2_{\rm tot}$, $\sigma^2_{\rm for}$ and
$\sigma^2_{\rm det}$ deviate by no more than 5$\%$ from their minimum values
(top right, bottom right and bottom left panel
respectively).
Results expected from SWIPE only (red) are also shown for comparison.

\subsection{Including $E$--modes}
\label{Emodes}

In this section we investigate the effect of including $E$--modes
in the analysis. Since $E$--modes should provide strong constraints
on the optical depth $\tau$,
as explained in section \ref{FMforecasts}, we extend the set of parameters
to be explored in FM evaluations by adding $\tau$ and $A_s$, namely
${\bf p} =\{ r,\tau,\ln(10^{10} A_s),{\cal A}_S^B,{\cal A}_D^B,\beta_S,\beta_D \}$,
and we assume ${\cal A}_{S(D)}^B=0.34(0.56){\cal A}_{S(D)}^E$ according to \cite{planckX}.
Results are summarized in Figure \ref{sn} which shows the behavior of
the signal to noise ratio $r/\sigma_r$ as a function of $r$ in the following
cases satisfying FoM 1:
\begin{itemize}[noitemsep]
\item  SWIPE, $B$--modes only
\item  SWIPE, $B$-- and $E$--modes
\item  SWIPE+STRIP, $B$-- and $E$--modes
\end{itemize}
For SWIPE (STRIP) we consider both the combinations  $90+220+240$ GHz
and $140+220+240$ GHz ($43$ GHz only ($n=0$) and $43+90$ GHz ($n=5$)).

As already observed in previous sections the presence of the 90 GHz high
frequency channel substantially improves the precision with which $r$
is measured.
This is further enhanced by adding low frequency channels and improves
as more 43 GHz modules are replaced by 90 GHz ones. The
STRIP($43+90$ GHz)+SWIPE($90+140+220$ GHz) combination would then  be able
to measure $r \simeq 0.06$ at $3-\sigma$ level and set upper limits if
$r\lesssim 0.02$. On the other hand, if replacing the SWIPE($90+140+220$ GHz)
with the SWIPE($140+220+240$ GHz), a $3-\sigma$ detection is only possible
for $r\simeq 0.1$. 
Finally, in Table \ref{sigtau} we give the uncertainties on $\tau$ and 
$\ln(10^{10}A_s)$ which turn out to be competitive with the current ones, 
these latter being $\sigma_\tau=0.017$ and
$\sigma_{\ln(10^{10}A_s)}=0.034$ \cite{planckXIII2015}.

\begin{table}
\begin{center}
\renewcommand\arraystretch{1.2}
\begin{tabular}{|ccc|}
\hline
\multicolumn{3}{|c|}{SWIPE 140+220+240 GHz ($E+B$)} \\
\hline
  & $\sigma_\tau$ & $\sigma_{\ln(10^{10}A_s})$ \\
SWIPE & 0.023 & 0.054\\
STRIP+SWIPE ($n=0$) & 0.022 & 0.051 \\
STRIP+SWIPE ($n=5$) & 0.014 & 0.039\\
\hline
\multicolumn{3}{c}{} \\
\hline
\multicolumn{3}{|c|}{SWIPE 90+140+220 GHz ($E+B$)} \\
\hline
  & $\sigma_\tau$ & $\sigma_{\ln(10^{10}A_s})$ \\
SWIPE & 0.019 & 0.046\\
STRIP+SWIPE ($n=0$) & 0.013 & 0.036 \\
STRIP+SWIPE ($n=5$) & 0.013 & 0.036 \\
\hline
\end{tabular}
\caption{Uncertainties on $\tau$ and  $\ln(10^{10}A_s)$ for different cases considered
in the text. Note that $\sigma_\tau$ and $\sigma_{\ln(10^{10}A_s})$ do not depend on the value
of $r$.}
\label{sigtau}\end{center}
\end{table}

\section{Conclusions}
\label{conclusions}

In this work we have performed optimization and forecasting analisys for
LSPE experiment aimed to CMB $B$--modes measurements.
To this aim we have investigated various configurations
of the focal plane obtained by considering different combinations of
frequency channels and/or varing the number of detectors within each channel
(keeping their total number fixed).
Four Figures of Merit (FoM), based on the  forecasted uncertainty $\sigma_r$ of
the tensor--to-scalar ratio $r$ and the expected map variance from
foreground and instrumental noise residuals, were then defined in order
to identify optimal focal planes (see Section \ref{high}).

For the balloon--borne high frequency instrument SWIPE, we considered
a combination of 3 frequencies, chosen among 90, 140, 220 and 240 GHz.
On the other hand, the low frequency one, STRIP, was assumed to consist
of 2 channels at 43 and 90 GHz.

We found that the presence of the high frequency 90 GHz channel substantially
improve the precision on $r$ (by a $40-60 \%$ depending on the details of
the configurations).
Nevertheless, given the impossibility to
accomodate large size detectors into the focal plane and the aperture
constraints for the SWIPE telescope, such 90 GHz detectors would not
comply with the baseline requirement of maintaining a higher number of
modes per detector, which was originally chosen for the focal plane design of
SWIPE.
Therefore, the 140+220+240 GHz high frequency combination turns out to be
the preferred one. In this case, however, the configuration which provides 
the higher signal to noise $r/\sigma_r$ is quite different
from the one yielding the lower variance from non--cosmic signal
(see e.g. Table \ref{confrontoval}
and Figure \ref{confronto}, FoM 1 compared to FoM 2 and 3) so that a possible
compromise between the two should be considered.

Constraints on $r$ are somewhat improved if high frequency measurements
are complemented by 43 GHz low frequency ones and are further enhanced
as more as 43 GHz detectors are replaced by 90 GHz (low frequency) ones.
Effects of reducing the instrumental noise (e.g. by increasing the
observation time) are also discussed in Section \ref{noisered}.

We, however, advise the reader that, because of the issues and limitations
(described at the end of Section \ref {FMforecasts}) due
to our simplified procedure,
our findings should be considered as indicative
of the need for more investigation to be performed through realistic
simulations of the sky to which apply foreground separation/cleaning procedures.

Although only indicative, our results could however be of some interest
for the design of LSPE--like experiments.

\appendix
\section{Foreground removal}
\label{foregroundremoval}

In order to perform our analisys and include
the effect of foregrounds in the Fisher Matrix formalism,
we follow the method described in \cite{tegmark} which permits
to minimize the non--cosmic signal (instrumental noise and
foreground contaminations) by taking a suitable linear combination of
sky maps observed at different frequencies.
For each focal plane configuration, the final result is two
cleaned polarization maps ($E$-- and $B$--modes) which are
unbiased and have the smallest rms errors from instrumental noise
and foregrounds combined. Here below, we briefly review the main results
of the procedure while
refer the reader to \cite{tegmark} for more technical details.

Let us suppose a multifrequency
experiment measuring  $E$ and $B$ and let $n_{\nu}^X$ be
the number of frequency channels with which $X=E,B$ is observed.
The spherical harmonic coefficients observed at the frequency $\nu$
will then be:
\begin{equation}
\label{alm}
a_{lm}^{X\nu}=c_{lm}^{X}+ r_{lm}^{X\nu} 
\end{equation}
where $c_{lm}^{X}$ is the (frequency independent) CMB signal and
$ r_{lm}^{X\nu} = n_{lm}^{X\nu} + \sum_j f_{lm,j}^{X\nu}$
includes the instrumental noise $n_{lm}^{X\nu}$ and the possible contribution
of $j$ different foregrounds $f_{lm,j}^{X\nu}$.
In vector notation we can write:
$$
{\pmb a}_{lm} = \left[
\begin{array}{c}
{\pmb a}^E_{lm}\\
{\pmb a}^B_{lm}\\
\end{array}
\right]
\qquad
{\pmb c}_{lm} = {\pmb A} \left[
\begin{array}{c}
 c^E_{lm}\\
 c^B_{lm}\\
\end{array}
\right]
\qquad
{\pmb r}_{lm} = \left[
\begin{array}{c}
{\pmb r}^E_{lm}\\
{\pmb r}^B_{lm}\\
\end{array}
\right]
$$
\begin{equation}
{\pmb a}_{lm} = {\pmb c}_{lm}+{\pmb r}_{lm}
\end{equation}
where ${\pmb a}_{lm}$, ${\pmb c}_{lm}$ and ${\pmb r}_{lm}$ are vectors of length
$n_{\nu}=n_{\nu}^E+n_{\nu}^B$, the entries of
${\pmb a}^X_{lm}$ (${\pmb r}^X_{lm}$) are the $n_{\nu}^X$ measured
harmonic coefficients $a_{lm}^{X\nu}$ ($r_{lm}^{X\nu}$) given by eq.
(\ref{alm}) and the $n_\nu \times 2$ scan matrix ${\pmb A}$ is given by:
\begin{equation}
{\pmb A}=\left[
\begin{array}{cc}
 {\pmb e}^E & {\bf 0}\\
{\bf 0}   & {\pmb e}^B
\end{array}
\right]
\end{equation}
(${\pmb e}^X$ being the $n_{\nu}^X$--dimensional column vector
consisting of ones).

We can then define the angular auto-- and cross--power
spectra  of CMB polarization,  detector noise and foregrounds:
\begin{eqnarray}
\nonumber
\langle { c_{lm}^{X \, \ast}} \;  c_{l^\prime m^\prime}^{X^\prime} \rangle
&=& C_l^{X} \delta_{ll^\prime}\delta_{mm^\prime}\delta_{XX^\prime} \\
\nonumber
\langle {n_{lm}^{X\nu \, \ast}}  \;  n_{l^\prime m^\prime}^{X^\prime\nu^\prime} \rangle
&=& N_l^{X \nu\nu} \delta_{ll^\prime}\delta_{mm^\prime}
\delta_{XX^\prime}\delta_{\nu\nu^\prime} \\
\langle {f_{lmj}^{X\nu \, \ast}} \;   f_{l^\prime m^\prime j^\prime}^{X^\prime\nu^\prime} \rangle
&=& F_{lj}^{X \nu \nu^\prime} \delta_{ll^\prime}\delta_{mm^\prime}\delta_{XX^\prime}
\delta_{jj^\prime}
\end{eqnarray}
and the $n_{\nu} \times n_{\nu}$ total
covariance matrix ${\pmb R}_l$ of the non--cosmic signal
(detector noise and foregrounds):
\begin{equation}
{\pmb R}_l = \left[
\begin{array}{cc}
 {\pmb R}^{E}_l    & {\bf 0}\\
 {\bf 0}        & {\pmb R}^{B}_l
\end{array}
\right]
\end{equation}
built by the
two $n_{\nu}^X \times n_{\nu}^X$ covariance matrices
${\pmb R}_l^{X}$
specifying the correlation of the
 non--cosmic signal among different
frequency channels. Their components read:
\begin{equation}
\nonumber \left[ {\pmb R}_l^{X \;}\right]_{ab\;} =
\left[ {\pmb N}_l^{X \;}\right]_{ab\;}+
\left[ {\pmb F}_l^{X \;}\right]_{ab\;}=
N_l^{X  \nu_a\nu_a}\delta_{ab}
+ \sum_j F_{lj}^{X  \nu_a\nu_b}
\end{equation}
(here $a,b=1,...,n_\nu^X$ while $\nu_a$ denotes the $a$--th
frequency at which $X$ is observed; ${\pmb N}_l^{X \;}$ and ${\pmb F}_l^{X \;}$
are the covariance matrices of detector noise and foregrounds respectively).
Note that, here above, we have assumed that:
i) CMB, instrumental noise and different
foregrounds are all uncorrelated with each other, i.e.
$\langle { c_{lm}^{X\nu \, \ast}} \;  n_{l^\prime m^\prime}^{X^\prime\nu^\prime}\rangle =
\langle { c_{lm}^{X\nu \, \ast}} \; f_{l^\prime m^\prime,j}^{X^\prime\nu^\prime}\rangle =
\langle { f_{lm,j}^{X\nu \, \ast}} \; n_{l^\prime m^\prime}^{X^\prime\nu^\prime}\rangle = 0$;
ii) the vanishing of the cross--correlations between $B$--modes and
 $E$--modes due to symmetry reasons holds also for foregrounds;
iii) as usual, instrumental noise
introduces no cross--correlation among  $E$ and $B$ as well as among
different frequency channels.

Let us now consider the linear combination
${\bf a}_{lm}= {\pmb W}_l^t {\pmb a}_{lm}$
($t$ denoting the transpose matrix) of sky maps observed at different
frequencies. If we require that
${\pmb W}_l^t {\pmb A} ={\bf I}$ the two resulting maps ($E$,$B$) may be written as:
\begin{equation}
{\bf a}_{lm}= {\bf c}_{lm} + {\bf r}_{lm}
\end{equation}
where:
\begin{equation}
\label{almcleaned}
{\bf a}_{lm}=
\left[
\begin{array}{c}
{\rm a}^E_{lm}\\
{\rm a}^B_{lm}
\end{array}
\right]
\qquad
{\bf c}_{lm}=
\left[
\begin{array}{c}
c^E_{lm}\\
c^B_{lm}
\end{array}
\right]
\qquad
{\bf r}_{lm}=
{\pmb W}_l^t {\pmb r}_{lm}=
\left[
\begin{array}{c}
{\rm r}^E_{lm}\\
{\rm r}^B_{lm}
\end{array}
\right]
\end{equation}
Thus, since the CMB signal does not depend on the frequency, it is
untouched regardless of the chosen ${\pmb W}_l$, while this latter may
be chosen to suppress the impact of the non--cosmic signal.
The variance of the resulting maps is then
minimized provided that the $n_\nu \times 2$ weighting matrix
${\pmb W}_l$ has the form (see \cite{tegmark}):
\begin{equation}
\label{Wl}
{\pmb W}_l= {\pmb R}_l^{-1}{\pmb A}\left({\pmb A}^t {\pmb R}_l^{-1}
{\pmb A}\right)^{-1}
\end{equation}

Eqs. (\ref{almcleaned}) and (\ref{Wl}) thus gives the spherical harmonic
coefficients of the two cleaned maps. In particular, ${\bf r}_{lm}$
gives the residual non--cosmic signal after the cleaning procedure.
Its ($2\times 2$) covariance matrix:
\begin{equation}
{{\rm \bf R}_{l}}= \left( {\pmb A}^t {\pmb R}_l^{-1} {\pmb A}\right)^{-1} =
{\pmb W}_l^t {\pmb R}_l {\pmb W}_l = \left[
\begin{array}{cc}
{\rm R}^{E}_l & 0 \\
0        & {\rm R}^{B}_l
\end{array}
\right]
\end{equation}
provides the angular power spectra
${\rm R}^{X}_l={\rm N}^{X}_l+{\rm F}^{X}_l$ of
residual detector noise and foregrounds in the cleaned maps
needed to implement the Fisher matrix approach of Sec. \ref{FMforecasts}
(see eq. (\ref{spectra})).

\section{Foreground models}
\label{foremodels}

In order to implement our forecasting method, foreground angular
power spectra should be properly modelled and, in this respect, we rely
on recent measurements provided by the Planck Collaboration
\cite{planckXXII,planckX,planckXXX}.

Of all diffuse galactic foregrounds, dust ($D$) and synchroton ($S$) are
the relevant ones for polarization measurements.
Angular power spectra for $j=D,S$ are modeled as:
\begin{equation}
F_{lj}^{X \nu \nu^\prime} = {\cal A}^{X}_j \Theta_{lj}^{X}  {\cal F}^{\nu}_j {\cal  F}^{\nu^\prime}_j {\cal R}^{\nu \nu^\prime}_j
\end{equation}
 Here, ${\cal A}^{X}_j$ is the overall amplitude
and spectra have been factorized into
a multipole dependence term $\Theta_{lj}^{X}$, a frequency dependence
term ${\cal  F}^{\nu}_j$ and a frequency correlation term
${\cal R}^{\nu \nu^\prime}_j$ which,
following \cite{tegmark98,tegmark}, we assume to be modelled as:
\begin{equation}
{\cal R}^{\nu \nu^\prime}_j \simeq  ~e^{-\frac{1}{2}\left[\frac{\ln(\nu/\nu\prime)}
{\xi_j}\right]^2}
\end{equation}
where the {\it frequency coherence} $\xi_j$ accounts for slight variations
of the frequency dependence of foregrounds across the sky. It determines how
many powers of $e$ we can change the frequency before the correlation
between the channels  starts to break down.
The two limits
$\xi \rightarrow 0 $ and $\xi  \rightarrow \infty$
correspod to the two extreme cases of no correlation and perfect correlation
respectively.
Real foregrounds tipically have behaviours that are intermediate between these
two. In  \cite{tegmark98} it is shown that for a spectrum
$\propto f(\nu)\nu^\beta$ ($f(\nu)$ being some arbitrary function)
we have $\xi \approx 1/\sqrt2 \Delta\beta$ where $\Delta\beta$ is
the rms dispersion across the sky of the spectral index $\beta$.
For simplicity we assume as fiducial value $\Delta\beta=0.2$
for both $S$ and $D$ according to the findings by \cite{planckXXII,fuskeland,kra}.

Specifications of foreground models are summarized in Table
\ref{foretab}.
Galactic foregrounds amplitudes, ${\cal A}^X_{D(S)}$, are the mean values
evaluated by Planck \cite{planckX} with $2^\circ$ FWHM apodization over
an effective sky fraction $f_{sky}=0.68$.
We assume these values to hold also in the region oberserved by LSPE.
Finally, FM  is evaluated by adopting the following Gaussian priors:
$$
\frac{\sigma_P({\cal A}^{B}_S)}{{\cal A}^{B}_S}=0.18 ~ , \qquad
\sigma_P(\beta_S)=0.04 ~ , \qquad
\frac{\sigma_P({\cal A}^{B}_D)}{{\cal A}^{B}_D}=0.02 ~ , \qquad
\sigma_P(\beta_D)=0.02
$$
which correspond to $\approx 1-\sigma$ errors as reported by 
\cite{planckXXII,fuskeland,planckX}.

\begin{table}
\begin{center}
\renewcommand\arraystretch{1.8}
 \begin{tabular}{|lll|}
\hline
\multicolumn{3}{|c|}{Synchrotron}\\
\hline
${\cal A}^{E}_S=0.0031 \; \mu {\rm K} \qquad $ &
$\Theta_{lD}^{X}= \left( \frac{l}{l_0} \right)^{\alpha^{X}} \qquad $ &
${\cal F}^{\nu}_S = \left( \frac{\nu}{\nu_0} \right)^{\beta_S} \frac{g_{\nu}}{g_{\nu_0}} $  \\
${\cal A}^{B}_S= 0.35 {\cal A}^{E}_S$ &
$\alpha^{E}=-2.49 $ &
$\beta_S=-3.14$\\
  &
$ \alpha^{B}=-2.32 $&
$\nu_0=30 \; {\rm GHz}$\\
 & $l_0=80$ & \\
\hline
\multicolumn{3}{|c|}{Thermal Dust}\\
\hline
${\cal A}^{E}_D=0.2765 \; \mu {\rm K} $ &
$\Theta_{lD}^{X}= \left( \frac{l}{l_0} \right)^{\alpha^{X}}  $ &
${\cal F}^{\nu}_D = \left( \frac{\nu}{\nu_0} \right)^{\beta_D+1} \frac{B_{\nu_0}}{B_\nu} \frac{g_{\nu}}{g_{\nu_0}} $  \\
${\cal A}^{B}_D= 0.56 {\cal A}^{E}_D$ &
$\alpha^{E}=-2.53 $ &
$\beta_D=1.59$\\
  &
$ \alpha^{B}=-2.62 $&
$\nu_0=353 \; {\rm GHz}$\\
 & $l_0=80$ & $B_\nu=e^{\frac{h\nu}{kT_D}}-1$ \\
&
 &
$T_D=19.6 \; {\rm K}$
\\
\hline
\end{tabular}
\caption{Specifications of foreground model. Here $g_\nu=(e^x-1)^2/x^2e^x$,
where $x=h\nu/k T_{CMB}$, accounts for the conversion from antenna to thermodynamic
temperature.}
\label{foretab}
\end{center}
\end{table}

\section{Instrumental noise}
\label{noise}

Assuming the instrumental noise is white and Gaussian, 
the noise angular power spectrum for $X=E,B$ reads:
\begin{equation}
N_l^{X\nu}=\sigma_{pix}^2 \, \Omega_{pix}
\, e^{l(l+1)\sigma_B^2}
\label{noisespec}
\end{equation}
where $\sigma_{pix}$ and $\Omega_{pix}=\theta^2_{FWHM}$ are the pixel
noise and area respectively, $\theta_{FWHM}$ is the full width at half
maximum beamsize and $\sigma_B=\theta_{FWHM} /\sqrt{8\ln2}$.
These quantities depends on the features of the detectors and are summarized
in Table \ref{noisetab}.

Note that, for simplicity, we have assumed 
the detectors to have Gaussian beam profiles that is not the case with the
multimode detectors of SWIPE. 
Sensitivities in Table \ref{noisetab} are however calculated on the base 
of realistic multimode throughputs.

\begin{table}
\begin{center}
\renewcommand\arraystretch{1.2}
\begin{tabular}{|l|cc|cccc|}
\hline
& \multicolumn{2}{|c|}{STRIP} & \multicolumn{4}{|c|}{SWIPE} \\
\hline
$\nu/$GHz& $43$ & $90$ & $90$ & $140$ & $220$ & $240$ \\
$\theta_{FWHM} /$arcmin
& $30.00$ & $15.00$  $^b$& $85.95$ & $85.95$ & $85.95$ &  $85.95$\\
$\sigma_{pix}/\mu$K $\quad E,B$ $^a$
 &  $7.84$ & $8.78$ &  $2.10$ &  $1.35$ &  $5.51$ & $11.47$    \\
\hline
Observing time & \multicolumn{2}{|c|}{1 year} & \multicolumn{4}{|c|}{14 days}\\
Duty cycle (\%)& \multicolumn{2}{|c|}{35} & \multicolumn{4}{|c|}{100}\\
$f_{sky}$ &  \multicolumn{2}{|c|}{0.20} &  \multicolumn{4}{|c|}{0.20} \\
\hline
 \multicolumn{7}{l}{$^a$ $\sigma_{pix}$ is the sensitivity per $1^\circ$ pixel per detector}\\
 \multicolumn{7}{l}{$^b$ $\theta_{FWHM}=30$ arcmin  for the 6 detectors aimed
at monitoring the atmospheric emission (see text)}
\end{tabular}
\caption{Characteristics of LSPE instruments.}
\label{noisetab}
\end{center}
\end{table}

\section{Angular power spectra covariance matrix}
\label{covariance}

Given the binned angular power spectra ${\rm D}_B^X$ ($X=E,B$)
as defined in (\ref{binnedspectra})
and (\ref{spectra}):
\begin{equation}
{\rm D}_B^X= \frac{1}{\Delta l}\sum_{l\in b}{\rm D}_l^X \qquad {\rm D}_l^X=\frac{l(l+1)}{2\pi}\left( C_l^X
+ {\rm R}_l^X \right)
\end{equation}
their covariance matrix  ${\bf D}_b$ reads:
\begin{equation}
{\bf D}_b=\sum_{l\in b} \frac{2}{(2l+1)f_{sky} \Delta l^2}
\left[
\begin{array}{cc}
 \left({\rm D}_l^{E}\right) ^2  & 0 \\
 0  &  \left({\rm D}_l^{B}\right) ^2  \\
\end{array}
\right]
\end{equation}
where $f_{sky}$
is the observed sky fraction.

\acknowledgments
Authors wish to thank Paolo de Bernardis for many useful insights. This work is supported by ASI (Agenzia Spaziale Italiana)
under contract ASI I/022/11/0 "Large Scale Polarization Explorer (LSPE)".

\end{document}